\documentclass[twocolumn,english,aps,floatfix,superscriptaddress,showpacs,amsfonts,amssymb,superscriptaddress]{revtex4}
\usepackage{color}

\usepackage{graphicx}
\usepackage{amsmath}
\usepackage{latexsym}
\usepackage{epstopdf}
\usepackage{amsmath}
\usepackage{amssymb,amsmath}
\usepackage{multirow}
\usepackage{mathtools}
\usepackage{tikz}
\usetikzlibrary{patterns}

\newcommand{\beq}{\begin{equation}}
\newcommand{\eeq}{\end{equation}}

\def\bra#1{\langle #1|}
\def\ket#1{|#1\rangle}

\newcommand{\eq}{\begin{equation}}
\newcommand{\en}{\end{equation}}
\newcommand{\ear}{\begin{eqnarray}}
\newcommand{\rae}{\end{eqnarray}}

\newcommand{\tr}{{\rm tr}\,}

\def\bra#1{\langle #1|}
\def\ket#1{|#1\rangle}

\begin{document}
\title{Formation probabilities and Shannon information and their time evolution after quantum quench in transverse-field XY-chain}

\author{Khadijeh Najafi }
\affiliation{ Department of Physics, Georgetown University, 37th and O Sts. NW, Washington, DC 20057, USA}
\author{ M.~A.~Rajabpour}
\affiliation{ Instituto de F\'isica, Universidade Federal Fluminense, Av. Gal. Milton Tavares de Souza s/n, Gragoat\'a, 24210-346, Niter\'oi, RJ, Brazil}

\date{\today{}}

\begin{abstract}
We first provide a formula  to calculate  the probability of occurrence of different configurations
(formation probabilities) in a generic free fermion system.
We then study the scaling of these probabilities with respect to the size in the case of critical 
transverse-field XY-chain in the $\sigma^z$ bases. In the case of the transverse field Ising model, we show that all
the "crystal" configurations follow the formulas expected from conformal field theory (CFT). In the case of critical $XX$ chain, we show that
the only configurations that follow the formulas of the CFT are the ones which respect the filling factor of the system. By repeating
all the
calculations  in the presence of open and periodic boundary conditions we find  further support to our classification of different configurations.
Using the developed  technique, we also study Shannon information of a subregion in our system.  In this respect we 
distinguish particular configurations that are more important in the study of the scaling limit of the Shannon information of the subsystem. Finally, we 
study the evolution of formation probabilities,
Shannon information and Shannon mutual information after a quantum quench in free fermion system. In particular,
for the intial state considered in this paper, we demonstrate that  
the  Shannon information  after quantum quench first increases with the time and then saturates at time $t^*=\frac{l}{2}$, where $l$
is the size of the subsystem.

\end{abstract}
\maketitle
\section{Introduction}

Studying correlation functions in many-body systems has been considered one of the main topics in statistical mechanics and condensed matter
physics for many years. Although, for a long-time the main quantities of interest were the correlation functions of local observables
the recent interest in calculating non-local quantities, especially the  entanglement entropy, has made significant changes. One of the main reasons
for this interest (at least in $1+1$ dimensional critical systems) is that by calculating some of the non-local observables one can derive 
the central charge of the system
without referring to the velocity of sound, for the case of entanglement entropy see \cite{CC2009}. Another non-local
quantity which has been studied for many years with Bethe ansatz techniques and some other methods is emptiness formation probability 
\cite{Essler,U1,Shiroishi,Abanov-Korepin,Franchini,Stephan2013}.
In the case of spin chains, it is the probability of finding a sequence of up spins in the system (note that almost all of
the studies in this regard concentrate on the $\sigma^z$ bases). These studies show that this probability,
with respect to the sequence size, decreases  like a Gaussian in the case of systems with $U(1)$ symmetry \cite{U1} 
and exponentially in other cases \cite{Shiroishi,Franchini}. In the critical cases, the Gaussian and exponential are accompanied with
a power-law decay with a universal exponent. In a recent development \cite{Stephan2013} it was shown that
for those critical systems without $U(1)$ symmetry this universal exponent is dependent on the central charge of the system. The argument is
based on connecting the configuration of all spins up to some sort of boundary conformal field theory. One should notice that
the argument works just for those bases that can be connected to boundary conformal field theory. 

In an apparently connected
studies recently many authors investigated the Shannon information of quantum systems in different systems 
\cite{Stephan2009,Stephan2010,Oshikawa2010,Stephan2011,Alet2013,Alet2014a,Alet2014b,Alet2014c,Trebst2014,Trebst2015,Jose}. The Shannon information is
defined as follows: Consider the normalized ground state eigenfunction of a quantum spin chain Hamiltonian  
$\ket{\psi_G}=\sum_I a_I\ket{I}$, expressed in a particular local bases 
$\ket{I}=\ket{i_1,i_2,\cdots}$, where $i_1,i_2,\cdots$ are the eigenvalues of 
 some local operators  defined on the lattice sites. The Shannon entropy is defined as 
\begin{equation} \label{Shannon}
Sh =-\sum_I p_I \ln p_I,
\end{equation}
where $p_I=|a_I|^2$ is the probability of finding the system in  the 
particular configuration given by $\ket{I}$. As it is quite clear this quantity is bases dependent and to calculate it
one needs in principle to know the probability of occurrence of all the configurations. The number of all
configurations increases exponentially with the size of the system and that makes analytical and numerical
calculations of this quantity quite difficult. Note that the emptiness formation probability is just one of the whole possible
configurations. The Shannon information of the system changes like a volume law with respect to the size of the system so in
principle one does not expect to extract any universal information by studying the leading term. The universal quantities should come 
from the subleading terms. To study subleading terms, it is useful to define yet another quantity called Shannon mutual information. 
By considering   local bases it is always possible to
decompose the configurations as a combination of the configurations  
 inside and outside of a  subregion $A$ as $\ket{I}=\ket{I_AI_{\bar{A}}}$. Then one can define the marginal probabilities as
$p_{I_{A}}=\sum_{I_{\bar{A}}} p_{I_{A}I_{\bar{A}}}$ and 
$p_{I_{\bar{A}}}=\sum_{I_A} p_{I_AI_{\bar{A}}}$ for the subregion $A$ and its complement $\bar{A}$. 
Then the Shannon mutual information is 
\begin{equation} \label{Renyi2}
I(A,\bar{A})= Sh(A)+Sh(\bar{A})-Sh(A\cup\bar{A}),
\end{equation}
where $Sh(A)$ and $Sh(\bar{A})$ are the Shannon informations of the subregions $A$ and  $\bar{A}$.  From now on instead 
of using $p_{I_AI_{\bar{A}}}$
we will use just $p_I$. Since in the above quantity the volume part of the Shannon entropy disappears Shannon mutual information provides a useful
technique to study the subleading terms. Note that one can similarly define the above quantity also for two regions $A$ and $B$,
i.e. $I(A,B)$ that are not 
necessarily complement of each other.
The Shannon mutual information has been studied in many classical \cite{Wolf2008,wilms,Grassberger,Rittenberg2014} and quantum systems
\cite{Um,AR2013,Stephan2014,AR2014,AR2015}. Numerical studies on variety of different periodic quantum critical spin chains 
show that for particular bases (so called conformal bases) we have \cite{AR2013,AR2014,AR2015}
\begin{equation} \label{mutual}
I(A,\bar{A})= \beta \ln[\frac{L}{\pi}\sin(\frac{\pi l}{L})],
\end{equation}
where $L$ and $l$ are the total size and subsystem size respectively and   $\beta$ is  very close to $\frac{c}{4}$, with $c$  the central charge of 
the system.
 Note that  if one takes an arbitrary base (non-conformal bases) the coefficient $\beta$ is nothing to do with the central charge. It is worth mentioning
 that based on \cite{AR2014} the conformal bases are those bases that can be connected to some sort of boundary conformal field theory in the sense 
 of \cite{Stephan2013}. For example in the transverse field Ising model the $\sigma^x$ and $\sigma^z$ bases are the conformal bases. Note
 that if one considers the Shannon information of the subsystem we will have $Sh(A)=\alpha l+\beta \ln[\frac{L}{\pi}\sin(\frac{\pi l}{L})]$.
 Since
 to extract the Shannon information one needs to use all the probabilities the only way to consider all of them in the numerical
 calculations is exact diagonalization. This makes the numerical calculation for large sizes very difficult. The results of the articles
\cite{AR2013,AR2014,AR2015} are all for periodic systems with $L=30$ whenever there are spin one-half system and smaller sizes for systems with bigger spins. In a recent
work \cite{Stephan2014} the author was able to consider an infinite system and study the Shannon information of the subsystem 
up to the size $l=40$. It was concluded that for the $XX$ chain  $\beta$ is $\frac{c}{4}$ with $c=1$ but for the Ising model
although it is very close to $\frac{c}{4}$ with $c=\frac{1}{2}$ the results are suggesting that probably  $\beta$ is not exactly connected to
the central charge. Notice that all of the above calculations are done by considering periodic boundary conditions
for the connected regions $A$ and $\bar{A}$. It is not clear 
how the equation (\ref{mutual}) might change if one considers open boundary conditions. Finally it is worth mentioning that some of the 
the above results have recently been extended to disconnected regions in \cite{Rajabpour2015d}. 

Motivated by the studies of emptiness formation probability and Shannon information of the subsystem we study here some related quantities.
First of all, as it is natural one might be interested in studying the scaling limit of some other configurations with respect
to the size of the subsystem. For example, consider an antiferromagnetic configuration in the Ising model or any other configuration
with pattern. It is very important  to know that these configurations are also flowing to some sort of
boundary conformal field theory or not. This study will clearly also help to understand
the nature of the  Shannon mutual information. In addition, this kind of studies also are very useful in the calculations  
of post-measurement entanglement entropy and localizable entanglement entropy \cite{Rajabpour2015b,Rajabpour2015c}. Having the above motivations in mind, we study formation probabilities
, Shannon information and their evolution after a quantum quench in the quantum $XY$ chain in the $\sigma^z$ bases.

The outline of the paper is as follows: In the next section we will first define our system of interest, i.e. $XY$ chain and then we
will provide a method to calculate the probability of any configuration
in the free fermionic systems. In section three we will list all the known analytical results regarding emptiness formation probability
for infinite systems and also finite systems with periodic and open boundary conditions. Explicit distinguishment is made between
critical Ising model and $XX$ chain with $U(1)$ symmetry. In section IV, we will define many different configurations
with specially defined pattern and calculate their corresponding probabilities numerically. Here again, we discuss Ising and $XX$ universalities separately
for infinite systems and for the systems with boundary. We also discuss configurations that do not have any pattern.
In section V, we study the Shannon information in the transverse field critical Ising model and also critical $XX$ chain. We will classify
the configurations based on their magnetization and show that in principle just a small part of the configurations can have a finite contribution
to Shannon information in the scaling limit. Section VI is devoted to the evolution of formation probabilities
and Shannon information after a quantum quench. We prepare the system in a particular state and then we let it evolve with another Hamiltonian
and study the time evolution of the formation probabilities and especially Shannon information. Finally, the last section is about
our conclusions and possible future works.

\section{Formation probabilities from reduced density matrix}

In this section, we first define the system of interest and after that using the reduced density matrix of this system we will
find a very efficient method to calculate  formation probabilities for systems that can be mapped to free fermions.
The Hamiltonian of the XY-chain is  as follows:
\begin{eqnarray}\label{Hamiltonian Ising}
H=-\sum_{j=1}^L\Big{[}(\frac{1+a}{2})\sigma_j^x\sigma_{j+1}^x+(\frac{1-a}{2})\sigma_j^y\sigma_{j+1}^y+h\sigma_j^z\Big{]}.
\end{eqnarray}
After using  Jordan-Wigner transformation, i.e. 
$c_j=\prod_{_{m<j}}\sigma_m^z\frac{\sigma_j^x-i\sigma_j^y}{2}$ and $\mathcal{N}=\prod_{_{m=1}}^{^{L}}\sigma_m^z=\pm1$ 
with $c_{L+1}^{\dagger}=0$ and $c_{L+1}^{\dagger}=\mathcal{N}c_{1}^{\dagger}$ for open and periodic boundary conditions respectively
 the Hamiltonian will have the following form:
 \begin{widetext}
 \begin{equation}\label{Ising Hamiltonian free fermion}
H=\sum_{j=1}^L \Big{[}(c_j^{\dagger}c_{j+1}+ac_j^{\dagger}c_{j+1}^{\dagger}+h.c.)-h(2c_j^{\dagger}c_j-1)\Big{]}+
\mathcal{N}(c_L^{\dagger}c_1+ac_L^{\dagger}c_1^{\dagger}+h.c.).
\end{equation}
\end{widetext}

The above Hamiltonian has a very rich phase space with different critical regions \cite{Mccoy1971}. In figure ~1 we show different critical regions of the system.
To calculate probability of formations for different patterns we  first write the reduced density matrix of a block of spins $D$  by using block Green matrices. 
Following \cite{Peschela,Barthel2008} 
we first define the   operators
\begin{eqnarray}\label{New operators}
a_i=c_i^{\dagger}+c_i,\hspace{1cm}b_i=c_i^{\dagger}-c_i.	
\end{eqnarray}
The block Green matrix is defined as
\begin{eqnarray}\label{Green matrix}
G_{ij}=\tr[\rho_Db_ia_j].	
\end{eqnarray}
The elements of the Green matrix can be calculated  following  \cite{Lieb} and we will mention their explicit form for 
open and periodic boundary conditions later.

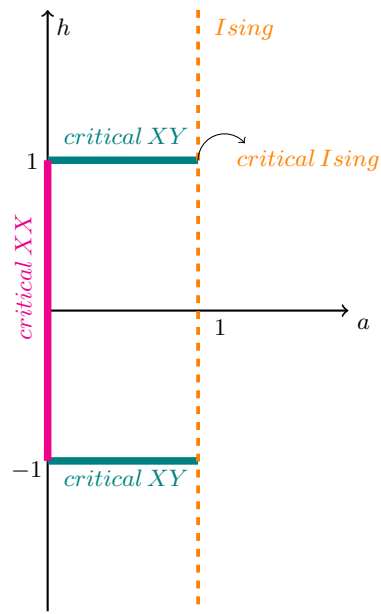
\begin{figure}
\begin{tikzpicture}[scale=2]
\draw [thick, ->] (0,0) -- (2,0);
\draw [thick] (0,-1) -- (0,-2);
\draw [thick, ->] (0,0) -- (0,2);
\draw [line width=0.1cm,teal] (0,1) -- (1,1);
\draw [line width=0.1cm,teal] (0,-1) -- (1,-1);
\draw [dashed,line width=0.05cm,orange] (1,2) -- (1,-2);
\draw [line width=0.1cm,magenta] (0,1) -- (0,-1);
\node [below right] at (2,0) {$a$};
\node [below right] at (0,2) {$h$};
\node [below right] at (-0.2,1.10) {$1$};
\node [below right] at (-0.3,-0.95) {$-1$};
\node [below right] at (1.05,0) {$1$};
\draw [->] (1,1) arc (180:40:5pt);
\node [above right,rotate=90, magenta] at (-.05,-0.25){$critical\, XX$};
\node [above right, teal] at (.05,1.05){$critical\, XY$};
\node [below right, teal] at (.05,-1.0){$critical\, XY$};
\node [below right, orange] at (1.05,2){$Ising$};
\node [right, orange] at (1.20,1){$critical\,Ising$};
 
\end{tikzpicture}
\caption{(Color online) Different critical regions in quantum $XY$ chain. The critical $XX$ chain has central charge $c=1$ and critical $XY$ chain has $c=\frac{1}{2}$.}
\end{figure}


To calculate the reduced density matrix after partial measurement we need to first define fermionic coherent states. They can be defined as follows

\begin{eqnarray}\label{fermionic coherent states1}
 |\boldsymbol{\xi}>= |\xi_1,\xi_2,...,\xi_N>= e^{-\sum_{i=1}^N\xi_ic_i^{\dagger}}|0>,
\end{eqnarray}
where $\xi_i$'s are Grassmann numbers following the  properties: $\xi_n\xi_m+\xi_m\xi_n=0$ and $\xi_n^2=\xi_m^2=0$. Then it is easy to show that
\begin{eqnarray}\label{fermionic coherent states2}
c_i |\boldsymbol{\xi}>= -\xi_i  |\boldsymbol{\xi}>.
\end{eqnarray}
Using the coherent states (\ref{fermionic coherent states1}) the reduced density matrix has the following form \cite{Barthel2008}
\begin{eqnarray}\label{reduced density matrix xi}
\rho_D(\boldsymbol{\xi},\boldsymbol{\xi'})&=&<\boldsymbol{\xi}|\rho_D|\boldsymbol{\xi'}>\nonumber\\
&=&\det\frac{1}{2}(1-G)e^{\frac{1}{2}(\boldsymbol{\xi}^*-\boldsymbol{\xi}')^T
F(\boldsymbol{\xi}^*+\boldsymbol{\xi}')},
\end{eqnarray}
where $F=(G+1)(1-G)^{-1}$. One can use the above formula to extract the formation probability for arbitrary configuration in the
$\sigma_z$ bases as follows: first of all
to extract the probability of particular configuration we need to look to the diagonal elements of $\rho_D(\boldsymbol{\xi},\boldsymbol{\xi'})$.
When the spin in the $\sigma^z$ direction is up in the fermionic representation it can be understood as the lack of a fermion in that site
which in the language of coherent states means that the corresponding $\xi$ is zero in the equation (\ref{reduced density matrix xi}). After putting some of
the $\xi$'s equal to zero one will have a new reduced density matrix in the coherent state basis with this constraint that some of the spins are fixed to be up. 
In other words in the equation (\ref{reduced density matrix xi}) instead of $F$ we will have $\tilde{F}$ which is a sub-matrix of the matrix $F$.
The elements of the new reduced density matrix will be $\tilde{\rho}_D(\boldsymbol{\eta},\boldsymbol{\eta'})$, where we put the $\xi$'s corresponding
to the sites filled with fermions equal to $\eta$.
To extract the probability 
of formation one just needs  to integrate over all the  $\eta$'s that correspond to the down spins. In other words after using formulas 
of the Grassmann Gaussian  integrals the formation probability
will have the following formula
\begin{eqnarray}\label{Formation probability}
P(\mathcal{C}_n)=\det[\frac{1}{2}(1-G)] M_{F}^{\mathcal{C}_n},
\end{eqnarray}
where $M_{F}^{\mathcal{C}_n}$ is the minor of the matrix $F$ corresponding to the configuration $\mathcal{C}_n$. Notice that we just need {\it{principal}} 
minors of the matrix $F$. Since the sum of all the principal minors of the matrix $F$ is equal to $\det(1+F)$ the normalization is ensured.
We have $ \binom {l} {k}$ number of rank-k minors for matrix $F$ with size $l$. Summing over the number of all principal 
minors, one can obtain $2^l$ which is the number
of all possible configurations. Configurations with the same minor rank have the same number of up spins and by increasing $k$ we actually
increase the number of down spins in the corresponding configuration. For example $k=0(l)$  is the case with all 
spins up (down) and sometimes it is called emptiness formation.  The above formula gives a very efficient way to calculate the  formation 
probabilities in numerical approach. Since, as we will show, all of these probabilities are exponentially small with respect to the size of
the subsystem it is much easier to work with the logarithm of them and define logarithmic  formation probabilities
\begin{eqnarray}\label{logarithmic Formation probability}
\Pi(\mathcal{C}_n)=-\ln P(\mathcal{C}_n).
\end{eqnarray}

All of the calculations done in this paper are based on the formula (\ref{Formation probability}) and are performed using Mathematica. 
In the next sections, we will study the  formation probabilities
for different configurations with the crystal order (pattern formation probabilities) with respect to the size of the subsystem for some particular 
critical regions of the system. 

\section{Emptiness formation probability: known results}

Before presenting our results, we first  review here the well-known facts regarding emptiness formation probability ($k=0$ and $l$). 
For reasons that will be clear
in the next section, we will call these two configurations $x=0$ and $1$ respectively. Using Fisher-Hartwig
theorem the emptiness
formation probability was already exhaustively studied in \cite{Franchini}. The results were generalized to arbitrary conformal critical systems
in \cite{Stephan2013}. Due to the $U(1)$ symmetry there is an important difference between the XX critical line and the Ising critical point. For this
reason, we report the results regarding these two possibilities separately. 

\subsection{ Critical Ising point}
We list here the results regarding the logarithmic emptiness formation probability at the  critical Ising point ($a=h=1$).

\paragraph{Configuration $\mathcal{C}_{x=0}$, i.e.  $(|\uparrow,\uparrow,...\uparrow>)$: }

This configuration corresponds to $k=0$ and has the highest probability and using the equation (\ref{Formation probability}) one can easily show that
 \begin{eqnarray}\label{all up}
P(x=0)=\det[\frac{1}{2}(1-G)]
\end{eqnarray}
The above formula is valid independent of the boundary conditions and the size of the system. For the infinite system 
at the critical Ising point the $G$ matrix has the following form:
 \begin{eqnarray}\label{G Ising infinite}
G_{ij}=-\frac{1}{\pi(i-j+1/2)}.
\end{eqnarray}
Since the above matrix is a Toeplitz matrix 
using Fisher-Hartwig conjecture  in \cite{Franchini}  it was shown that 
the logarithmic probability for this configuration changes with the subsystem size as follows:
 \begin{eqnarray}\label{main formula1}
\Pi(x=0)=\alpha l+\beta\ln l+\gamma\frac{\ln l}{l}+\mathcal{O}(1),
\end{eqnarray}
 where  $\beta=\frac{1}{16}=0.0625$, and $\alpha$ and $\gamma$ are some non-universal numbers. At the critical point of the Ising model these numbers are known
 $\alpha=\ln 2-2C/\pi=0.11002$, with $C$  the Catalan constant and  $\gamma=-\frac{1}{32\pi}=-0.00994$. The $\frac{\ln l}{l}$ term
 is the result of the paper \cite{Stephan2013}.
 Using conformal field theory techniques in \cite{Stephan2013} it was argued that for generic critical systems the
 coefficient of the logarithm should be $\beta=\frac{c}{8}$, where $c$ is the central charge of the critical system. For the Ising
 universality class $c=\frac{1}{2}$.
 
 When the size of the total system is finite $L$ depending on the form of the boundary conditions, periodic or open; $G$ has the following
 two forms
  \begin{eqnarray}\label{G Ising PBC }
G^P_{ij}&=&-\frac{1}{L\sin(\frac{\pi(i-j+1/2)}{L})},\\
\label{G Ising OBC }
G^O_{ij}&=&-\frac{1}{2L+1}\Big{(}\frac{1}{\sin(\frac{\pi(i-j+1/2)}{2L+1})}+\frac{1}{\sin(\frac{\pi(i+j+1/2)}{2L+1})}\Big{)}.\nonumber \\
\end{eqnarray}
 Notice that for $L\to\infty$ the first equation reduces to (\ref{G Ising infinite}) and the second equation will give the result for a semi-infinite chain.
 The results for emptiness formation probabilities for the above two cases are \cite{Stephan2013}

\begin{widetext}
\begin{eqnarray}
\label{efp Ising PBC up}
\Pi^P(x=0)&=&\alpha l+\beta\ln[\frac{L}{\pi}\sin\frac{\pi l}{L}]+\gamma\pi\cot(\frac{\pi l}{L})\frac{\ln l}{L}+\mathcal{O}(1),\\
\label{efp Ising OBC up}
\Pi^O(x=0)&=&\alpha l+\beta^o\ln[\frac{4L}{\pi}\frac{\tan^2\frac{\pi l}{2L}}{\sin\frac{\pi l}{L}}]+
\gamma^o\pi\frac{2-\cos(\frac{\pi l}{l})}{\sin\frac{\pi l}{l}}\frac{\ln l}{L}+\mathcal{O}(1),
\end{eqnarray}
\end{widetext}
where $\beta=\frac{c}{8}$ and $\beta^o=-\frac{c}{16}$. It is also conjectured that $\gamma=-\frac{c}{8\pi a}$ and $\gamma^o=\frac{c}{32\pi a}$.
The above two equations are derived using boundary conformal field theory techniques and in principle they are valid in
the Ising case because $x=0$ configuration is related to the free conformal boundary condition in the conformal Ising model.

\paragraph{Configuration $\mathcal{C}_{x=1}$, i.e.  $|\downarrow,\downarrow,...\downarrow>$: } 
 
 This case, which is also studied in \cite{Franchini} and \cite{Stephan2013}, corresponds to $k=l$ and has the lowest probability. One can easily show that
\begin{eqnarray}\label{all down}
P(x=1)=\det[\frac{1}{2}(1-G)]\det F=\det[\frac{1}{2}(1+G)].
\end{eqnarray}
For the infinite system it follows similar formula as (\ref{main formula1}) with also an extra $\nu\frac{(-1)^l}{\sqrt{l}}$ term, in other words,
 \begin{eqnarray}\label{main formula2}
\Pi(x=1)=\alpha l+\beta\ln l+\nu\frac{(-1)^l}{\sqrt{l}}+\gamma\frac{\ln l}{l}+\mathcal{O}(1),
\end{eqnarray} 
where $\beta=\frac{c}{8}$. At the critical Ising point $\alpha=\ln 2+2C/\pi=1.27626$ and $\nu=-0.21505$ and $\gamma$ is unknown. 
The oscillating $\nu$ term is mathematically explained by using generalized Fisher-Hartwig conjecture in \cite{Franchini}. 
To the best of our knowledge its presence  at the critical point has not been understood by physical arguments \cite{footnote1}. 
Our numerical results in the next section will show that the term is present whenever the parity of the
number of down spins changes with the subsystem size. The term is very important to be considered in numerical
calculations to get reliable results for $\beta$ which is 
 the universal and the most interesting term. 
 
 When the size of the system is finite depending on the type of the boundary conditions boundary changing operators can play an important role. 
 The following formulas are presented in \cite{Stephan2013}:
 \begin{eqnarray}\label{efp Ising PBC down}
\Pi^P(x=1)&=&\alpha l+\beta\ln[\frac{L}{\pi}\sin\frac{\pi l}{L}]+...,\\
\label{efp Ising OBC down}
\Pi^O(x=1)&=&\alpha l+\beta_1^o\ln[\frac{L}{\pi}\sin\frac{\pi l}{L}]+\beta_2^o\ln[\frac{L}{\pi}\tan\frac{\pi l}{2L}]+...,\nonumber \\
\end{eqnarray}
 where $\beta=\frac{c}{8}$, $\beta_1^o=\frac{c}{16}$ and $\beta_2^o=4h-\frac{c}{8}$ with $h=\frac{1}{16}$ being the conformal weight of
 the boundary changing operator. The dots are the subleading terms.
 
\subsection{ XX critical line} 

The critical XX chain $a=0$ has  $U(1)$ symmetry which as it is already discussed extensively in the literature  is the main reason for 
having Gaussian decaying emptiness formation probability \cite{U1,Stephan2013}. Since in this model $< c_i^{\dagger}c_j^{\dagger}>=< c_ic_j>=0$ 
the equation (\ref{reduced density matrix xi}) has simpler form
\begin{eqnarray}\label{reduced density matrix xx}
\rho_D(\boldsymbol{\xi},\boldsymbol{\xi'})=\det(1-C)e^{\boldsymbol{\xi}^*F\boldsymbol{\xi'}}
\end{eqnarray}
where $C_{ij}=< c_i^{\dagger}c_j>$ and $F=C(1-C)^{-1}$. Finally we have 
\begin{eqnarray}\label{Formation probability XX}
P(\mathcal{C}_n)=\det[1-C] M_{F}^{\mathcal{C}_n}
\end{eqnarray}
 The form of the  $C$ matrices in the periodic and open cases are \cite{Fagotti}:
\begin{widetext} 
 \begin{eqnarray}\label{G XX PBC }
C^P_{ij}&=&\frac{n_f}{\pi}\delta_{ij}+(1-\delta_{ij})\frac{\sin(n_f(i-j))}{L\sin(\frac{\pi(i-j)}{L})},\\
\label{G-XX-OBC}
C^O_{ij}&=&\Big{(}\frac{1}{2}-(\frac{L}{2(L+1)}-\frac{n'_f}{\pi})\Big{)}\delta_{ij}+(1-\delta_{ij})\frac{1}{2(L+1)}
\Big{(}\frac{\sin(n'_f(i-j))}{\sin(\frac{\pi(i-j)}{2L+2})}-\frac{\sin(n'_f(i+j))}{\sin(\frac{\pi(i+j)}{2L+2})}\Big{)},
\end{eqnarray}
\end{widetext}
where $n_f=\frac{\pi}{L}\Big{(}2 \lceil\frac{L}{2\pi}\arccos(-h) \rceil-1\Big{)}$ is the Fermi
momentum and $n'_f=\frac{\pi}{2(L+1)}\Big{(}1+2 	\lfloor\frac{(L+1)}{\pi}\arccos(-h)) 	\rfloor\Big{)}$ with 
$\lceil x\rceil (\lfloor x\rfloor)$ as the closest integer larger (smaller) than $x$.

The all spins up and down configurations do not lead to conformal boundary conditions and so
none of the equations that we mentioned in the last subsection are valid. However, using Widom conjecture it is already known that, 
see for example \cite{Shiroishi},  the probabilities for both $\mathcal{C}_{x=0}$ and $\mathcal{C}_{x=1}$ show Gaussian behavior. 
For systems with $U(1)$ symmetry one expects the
following behavior for logarithmic emptiness formation probability \cite{U1}:
 \begin{eqnarray}\label{EFP XX }
\Pi(x=0)=\alpha_2 l^2+\alpha l+\beta \ln l+\mathcal{O}(1),
\end{eqnarray}
where $\beta=\frac{1}{4}$ for critical XX chain.

\section{Logarithmic pattern formation probabilities }

In this section, we study the logarithmic pattern formation probability defined as $\Pi(\mathcal{C})=-\ln P(\mathcal{C})$ with respect to the size of
the subsystem. The easiest configurations to study are those that have some kind of {\it{crystal}} structure. Although everything is already known
and checked numerically for the emptiness formation probabilities we will also report the results concerning these cases as benchmarks. Here we 
introduce the configurations that we studied numerically. None of these configurations have been considered before in the literature. 

We study here the configurations with $k=\frac{l}{2},\frac{l}{3},\frac{l}{4},...,\frac{l}{10}$ with crystal pattern and we call them
configurations $x=\frac{1}{2},\frac{1}{3},\frac{1}{4},...,\frac{1}{10}$ and for some ranks we will study the two most basic
cases. For example we will study

\begin{description}

\item {\it{Configurations with}} $k=\frac{l}{2}$:

\begin{description}
 \item  \item[a]
$(|\downarrow,\uparrow,\downarrow,\uparrow,...>)$

 \item  \item[b]  $(|\downarrow,\downarrow,\uparrow,\uparrow,\downarrow,\downarrow,\uparrow,\uparrow,...>)$ 
\end{description}
\item {\it{Configurations with}} $k=\frac{l}{3}$:

\begin{description}
 \item  \item[a]
$(|\uparrow,\uparrow,\downarrow,\uparrow,\uparrow,\downarrow,\uparrow,\uparrow,...>)$

 \item  \item[b]
$(|\uparrow,\uparrow,\uparrow,\uparrow,\downarrow,\downarrow,\uparrow,\uparrow,\uparrow,\uparrow,\downarrow,\downarrow,...>)$ 

\end{description}

\end{description}
All the configurations with the same $k$ belong to the cases with an equal rank of the minor in the equation (\ref{Formation probability}). Note 
that in all of the upcoming numerical calculations in every step we increase 
the size of the subsystem with a number which is devidable to the length of the base of the corresponding configuration.
For
some particular $k$'s the $\textbf{a}$ and $\textbf{b}$ configurations differ by the parity effect. For example,
in $k=\frac{l}{2}a$  depending on   $l=4i$ or $l=4i-2$   with $i=1,2,...$ the subsystem has  even or odd
number of down spins. This means that the parity of the number of down spins changes with the subsystem size for this configuration.
However, for $k=\frac{l}{2}b$ this is not the case because in order to have "perfect crystal" in the subsystem
 we need to consider a subsystem with  $l=4i$ with $i=1,2,...$ which
has always even number of down spins inside. 
Because of this difference in parity effect for $k=\frac{l}{2}a$ and $k=\frac{l}{2}b$  we expect different subleading behavior for these two cases.
Finally notice that one can simply define configurations like $k=\frac{l}{2}c$ and $k=\frac{l}{3}c$ by simply taking bigger bases
for the crystals. For example, $k=\frac{l}{2}c$ can be understood as a configuration with the base: three  down spins and then three up spins.

\subsection{Transverse-field Ising chain}

Using the equation (\ref{Formation probability}) and (\ref{G Ising infinite}) we first studied the crystal configurations
introduced in the previous subsection for the case of infinite chain. To calculate the formation
probability for every configuration we first use the matrix $G$ introduced in (\ref{G Ising infinite}) to find the matrix $F$. Then for every
configuration we use an appropriate minor to calculate the corresponding probability in (\ref{Formation probability}). For example, 
in the case of $k=\frac{l}{2}a$
this can be done by just finding a minor of $F$ which can be derived by calculating the determinant of a submatrix $\tilde{F}$ obtained 
from $F$ by removing every other 
row and column.
The results for $\alpha$ and $\beta$ (the most interesting quantities in this study) are shown in the Table I. 
Based on the  numerical results  one can derive the following conclusions regarding crystal
configurations:
\begin{enumerate}
 \item All the crystal configurations follow either the equation (\ref{main formula1}) or (\ref{main formula2}) with
 $\beta=\frac{1}{16}$.
 \item Whenever the parity of the number of down  spins in a configuration changes with respect to the size of the subsystem
 we have the oscillating
 term $\frac{1}{\sqrt{l}}$. For example, in the case of $k=\frac{l}{2}a$ we have the subleading term $\frac{(-1)^{\frac{l}{2}}}{\sqrt{l}}$
 but $\frac{1}{\sqrt{l}}$ correction is absent in $k=\frac{l}{2}b$. It appears again for $k=\frac{l}{3}a$ in the form of $\frac{(-1)^{\frac{l}{3}}}{\sqrt{l}}$.
 Generalization to other configurations is strightforward.
 .

 \item Although in some cases $\alpha$ for bigger $x$ is smaller than $\alpha$ with smaller $x$ in average $\alpha$ increases with $x$.
\end{enumerate}

\begin{table}[hthp!]
\centering
{\begin{tabular}{|l|c|c|c|c|c|}
  \hline
      Configuration        & $\alpha$       & $\beta$  \\
      \hline
    $ x=0$                  & $0.110025$   & $0.062498$  \\ \hline
    $ x=1$                  & $1.276267$   & $0.062465$  \\ \hline
    $ x=\frac{1}{2} (a) $   & $0.984708$   & $0.062462$  \\ \hline
    $ x=\frac{1}{2} (b) $   & $0.755726$   & $0.062496$  \\ \hline
    $ x=\frac{1}{3} (a) $   & $0.818715$   & $0.062468$  \\ \hline 
    $ x=\frac{1}{3} (b) $   & $0.542109$   & $0.062491$  \\ \hline
    $ x=\frac{1}{4} (a) $   & $0.710620$   & $0.062481$  \\ \hline
    $ x=\frac{1}{4} (b) $   & $0.434286$   & $0.062524$  \\ \hline
    $ x=\frac{1}{5} (a) $   & $0.634016$   & $0.062495$  \\ \hline
    $ x=\frac{1}{6} (a) $   & $0.576551$   & $0.062509$  \\ \hline
    $ x=\frac{1}{7} (a) $   & $0.531651$   & $0.062523$  \\ \hline
    $ x=\frac{1}{8} (a) $   & $0.495482$   & $0.062537$  \\ \hline
    $ x=\frac{1}{9} (a) $   & $0.465643$   & $0.062549$  \\ \hline
    $ x=\frac{1}{10}(a) $   & $0.440555$   & $0.062562$  \\ \hline

   \hline
\end{tabular}}
   \caption{Fitting parameters for the logarithmic formation probabilities of 
   different crystal configurations of the critical Ising chain discussed in the text. All the data were extracted by fitting the
   data in the range $l\in(2000,2500)$ to $\alpha l+\beta\ln l+\gamma\frac{\ln l}{l}+\delta\frac{1}{l}+\eta$ for those cases
   that do not show parity effect and to $\alpha l+\beta\ln l+\gamma\frac{\ln l}{l}+\nu\frac{(-1)^{m}}{\sqrt{l}}+\delta\frac{1}{l}+\eta$ 
   (with suitable $m$) for
   those cases that show parity effect \cite{footnote2}. } 
\end{table}

\begin{figure} [hthp!] \label{fig2}
\centering
\includegraphics[width=0.4\textwidth,angle =-90]{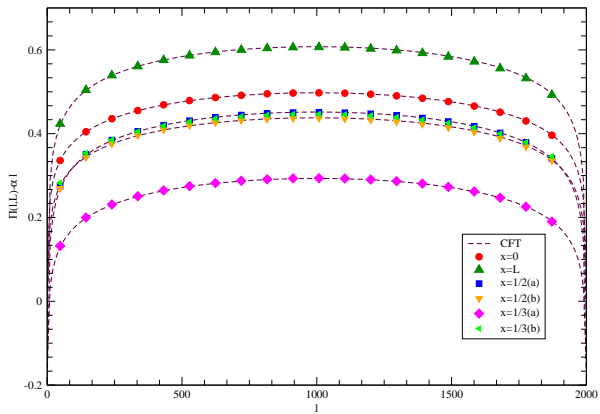}
\caption{(Color online) $\Pi(l,L)-\alpha l$ for periodic system with total length $L=2000$ with respect to $l$ for different configurations.
The dashed lines are the results expected from CFT, i.e. $\frac{1}{16}\ln [\frac{L}{\pi}\sin\frac{\pi l}{L}]+\eta$. } 
\end{figure}

\begin{figure} [hthp!] \label{fig3}
\centering
\includegraphics[width=0.4\textwidth,angle =-90]{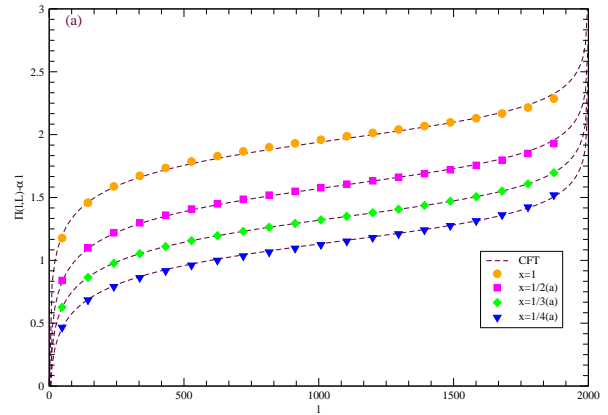}
\includegraphics[width=0.4\textwidth,angle =-90]{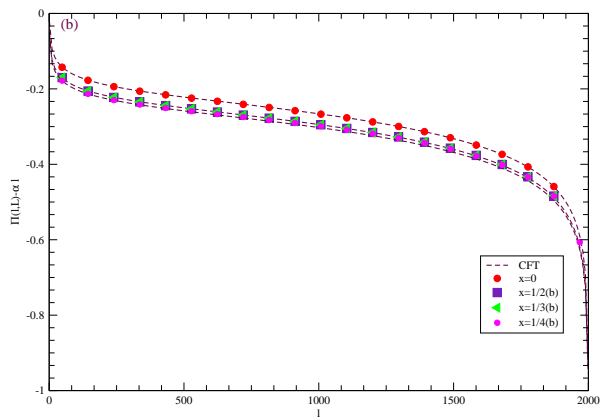}
\caption{(Color online) $\Pi(l,L)-\alpha l$ for open system with total length $L=2000$ with respect to $l$ for different configurations. a) configurations
without boundary changing operators and b) configurations with boundary changing operators.
The dashed lines are the results expected from CFT.} 
\end{figure}

We then studied the same configurations for the periodic  boundary condition. In the figure ~2  it is shown that
all of the configurations follow the formula (\ref{efp Ising PBC up}) . The case of the open boundary condition is more tricky
and depending on the configuration we have two possibilities: When the parity of the
number of down spins is  independent of the size of the subsystem
 (for example in the configurations $x=\frac{1}{2}b$ and $\frac{1}{3}b$)  we have the formula
(\ref{efp Ising OBC up}) but when we have the possibility of having odd or even number of down spins in a 
configurations (for example $x=\frac{1}{2}a,\frac{1}{3}a$) 
we have the formula (\ref{efp Ising OBC down}). The results are shown in the Figure ~3. This behavior could be anticipated based on the difference between
configurations $k=0$ and $k=l$ that we discussed before. It looks like that the boundary changing operator plays a role whenever there is
 the parity effect in the configuration. Looking to the problem in the language of Euclidean two dimensional classical system
 one can argue that in the case of open boundary condition we have a strip with a slit on it \cite{Stephan2013}. However, the boundary
 conditions on the boundary of strip can be different from the boundary condition on the slit, consequently, one needs to consider
 boundary changing operator on the point where the boundary condition changes. However, in general it is not clear which configurations
 lead to different boundary conditions on the slit and on the boundary of the strip. Our numerical results indeed 
 give a hint that depending on the bahavior of the  parity of the number of down spins in a configuration the conformal 
 boundary condition on the slit can be different. In the next two subsections, we will first comment on the validity of the above results in other cases 
 such as non-crystal configurations. Then we will also indicate the possible universal behavior of our results.
 
\subsubsection{Logarithmic formation probability of non-crystal configurations}

The number of crystal configurations is much smaller than the number of the whole configurations. In fact, the number of 
crystal configurations grows polynomially with the subsystem size but the number of whole configurations grows exponentially. However,
numerically it is very simple to check the formula for many configurations that have a small deviation from the crystal states.
For example, one can consider the case $k=1$ with all spins up except one and calculate the logarithmic formation 
probability using the equation (\ref{Formation probability}). It is clear that one does not expect the result be any different from
the equation (\ref{main formula1}) and indeed numerical results confirm this expectation. The important conclusion of this numerical
exercise is that there are many configurations "close" to crystal configurations that indeed follow either the equation 
(\ref{main formula1}) or equation (\ref{main formula2}) with all having the same $\beta$'s but different $\alpha$'s.

 The above results strongly suggest that all of the crystal and non-crystal configurations discussed in this section are flowing to some sort of
  conformal boundary conditions in the scaling limit. 

\subsubsection{Universality}

To check that the above results are the properties of the Ising universality class we also studied 
the critical XY-chain which has also central charge $c=\frac{1}{2}$. The Green matrix, in this case, is given by
 \begin{eqnarray}\label{Green matrix XY chain universality}
G_{ij}=\int_0^{\pi}\frac{d\phi}{\pi}\frac{(\cos\phi-1)\cos[(i-j)\phi]-a\sin\phi\sin\big[(i-j)\phi]}{\sqrt{(1-\cos\phi)^2+a^2\sin^2\phi}}.\nonumber
\end{eqnarray}
 Our numerical results depicted in the Figure ~4
show that the coefficient of the logarithmic term is a universal quantity which means that it has a fixed value on the critical XY-line.
The coefficient of the linear term changes by varying $a$ which indicates its non-universal nature.
 \begin{figure} [htb] \label{fig4}
\centering
\includegraphics[width=0.4\textwidth,angle =-90]{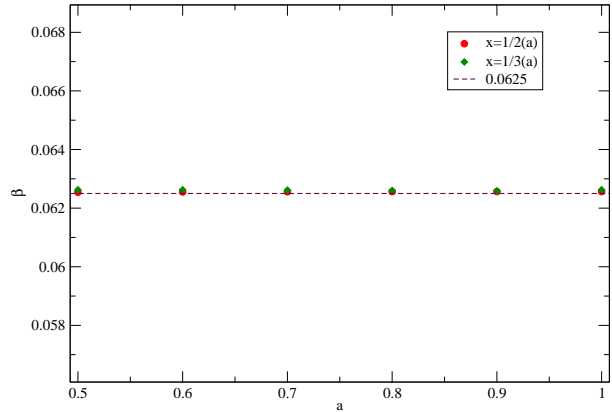}
\caption{(Color online) The coefficient of the logarithmic term in (\ref{main formula1}) for two configurations $x=\frac{1}{2}a$
and $x=\frac{1}{3}a$ for different values of $a$. The dashed line is the CFT result. The size of the largest subsystem was $l=500$ and all the results were extracted by fitting the
   data  to $\alpha l+\beta\ln l+\gamma\frac{\ln l}{l}+\nu\frac{(-1)^{m}}{\sqrt{l}}+\delta\frac{1}{l}+\eta$ 
   with suitable $m$ in the range $l\in(100,500)$. Estimated errors in the numerics are in the order of the size of the markers.} 
\end{figure}
\subsection{XX chain}

We repeated the calculations of the last section for also critical XX chain. The central charge of the system
is $c=1$. The results
of logarithmic formation probabilities for different magnetic field $h$ are shown in the Table II and Table III. 
 Based on the numerical calculations we conclude the followings:
  \begin{enumerate}
\item The configurations with $x=\frac{n_f}{\pi}$ follow  the equation (\ref{main formula1})  with
 $\beta=\frac{1}{8}$. This means that in the scaling limit most probably all of these configurations
 flow to some sort of  bounday conformal conditions. Note that as far as there is no boundary
 changing operator in the system the equation (\ref{main formula1}) is valid for any CFT independent 
 of its structure.
  \item All the other  configurations  follow the equation (\ref{EFP XX }) with $\beta$ which is different for different configurations.

\end{enumerate}
As we mentioned earlier XX chain has a  $U(1)$ symmetry which means that the number of particles is conserved.
The only configurations that respect this symmetry in the subsystem level
are the configurations with $x=\frac{n_f}{\pi}$. Any injection of the particles into the subsystem changes
drastically the formation probability. In the case of $x=0$ this phenomena is already explained in \cite{Stephan2013} based on arctic phenomena 
in the dimer model. It is quite natural to expect that similar structure is valid for all the configurations with $x\neq \frac{n_f}{\pi}$.
Note that based on our results the coefficient of the $\ln$ is $n_f$-dependent and strictly speaking is not a universal quantity.
\begin{table}[hthp!]
\centering
{\begin{tabular}{|c|l|c|c|c}
  \hline
      $n_{f}$     & Configuration        & $\alpha$       & $\beta$  \\ \hline
           &  $ x=1/2 (a) $   & $0.3465735$   & $0.124998$  \\ 
 $\pi/2$  & $ x=1/2 (b) $   & $0.5198604$   & $0.124997$  \\ 
           &  $ x=1/2 (c) $   & $0.7127780$   & $0.124597$  \\ \hline
 $\pi/3$ &  $ x=1/3 (a) $   & $0.3662041$   & $0.124987$  \\ \hline
 $\pi/4$ &  $ x=1/4 (a) $   & $0.3432345$   & $0.125024$  \\ \hline
 
   \hline
\end{tabular}}
   \caption{Fitting parameters for the logarithmic formation probability of antiferromagnetic configurations with different filling factors.
   All the results were extracted by fitting the data
in the range $l \in (100, 300)$ to $\alpha l+\beta\ln l+\gamma \frac{ \ln l}{l}+\delta\frac{1}{l}+\eta$.} 
\end{table}

\begin{table}[hthp!]
\centering
{\begin{tabular}{|l|c|c|c|c|c|}
  \hline
      Configuration         & $\alpha_{2}$   &  $\alpha$      &   $\beta$ \\
      \hline
    $ x=0,1$                & $0.346573$   & $0.000000$  &  $0.250054$   \\  \hline
    $ x=1/3 (a) $   & $0.035191$   & $0.366228$  &  $0.524293$   \\  \hline
    $ x=1/3 (b) $   & $0.035188$   & $0.597663$  &  $1.578683$   \\  \hline
    $ x=1/4 (a) $   & $0.080911$   & $0.346599$  &  $0.829767$   \\  \hline
    $ x=1/4 (b) $   & $0.080910$   & $0.587114$  &  $2.744492$   \\  \hline
    $ x=1/5 (a) $   & $0.118119$   & $0.321924$  &  $1.144949$   \\  \hline
    $ x=1/6 (a) $   & $0.147178$   & $0.298674$  &  $1.465399$   \\  \hline
    $ x=1/7 (a) $   & $0.170072$   & $0.278052$  &  $1.788319$   \\  \hline
    $ x=1/8 (a) $   & $0.188433$   & $0.260021$  &  $2.112155$   \\  \hline
    $ x=1/9 (a) $   & $0.203427$   & $0.244263$  &  $2.436061$   \\  \hline
    $ x=1/10(a) $   & $0.230432$   & $0.230432$  &  $2.759481$   \\  \hline

   \hline
\end{tabular}}  
\caption{Fitting parameters for different configurations with $x<\frac{1}{2}$ in the XX chain with $n_f=\frac{\pi}{2}$. All the data were extracted by fitting the data 
   in the range $l\in(100,300)$ to $\alpha_2 l^2+\alpha l+\beta \ln l+\eta$.  } 
\end{table}

We also studied the finite size effect in this model. The results of the numerical calculations
 of periodic boundary condition for $x=\frac{1}{2}$ are shown in the Figure ~5. It is shown that
 all of the configurations with $x=\frac{1}{2}$ follow the formula (\ref{efp Ising PBC up}) with $c=1$. 
 \begin{figure} [htb] \label{fig5}
\centering
\includegraphics[width=0.4\textwidth,angle =-90]{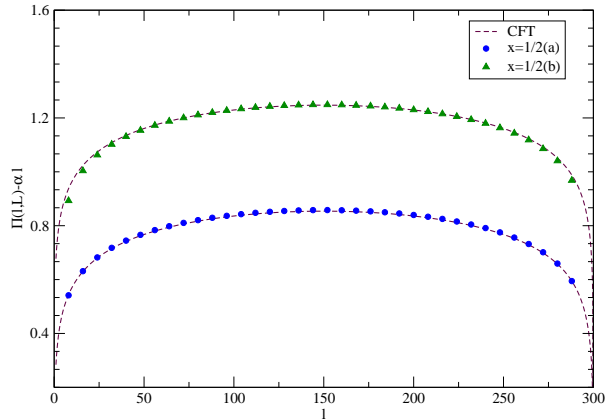}
\caption{(Color online) $\Pi(l,L)-\alpha l$ for periodic system with total length $L=300$ with respect to $l$ for different configurations
for critical $XX$-chain with $n_f=\frac{\pi}{2}$.
The dashed lines are the results expected from CFT.} 
\end{figure}

 We also repeated the same calculations for open
boundary conditions. We first performed the calculations for semi-infinite open chain and fitted the results to (\ref{main formula1})
and extracted the $\beta$. The coefficient of the logarithm not only depends on the rank of the configuration but also
to the configuration itself. It also changes with $n_f$. We were not able to find any universal feature in this case.

The above results  suggest that most probably all of the crystal configurations with $x=\frac{n_f}{\pi}$ in the periodic 
boundary condition flow to a boundary
conformal field theory. In the language of Luttinger liquid, the corresponding boundary condition should be the Dirichlet boundary condition \cite{Stephan2013}.
The case of the open boundary condition is intriguing and we leave it as an open problem.

\section{Shannon information of a subsystem }

In this section, we study Shannon information of a subsystem in transverse-field Ising model and $XX$ chain. For both models, the Shannon information is 
already calculated in \cite{Stephan2014} up to the size $l=40$ which it seems to be the current limit for classical computers. 
The reason that we are interested
in revisiting this quantity is to have a more detailed study of the contribution of different configurations. This will give an interesting insight regarding 
the possible scaling limit for this quantity. 

\subsection{Critical Ising }

In the last section, we studied many different configurations in the critical Ising model and we found that all of them follow
$P_{\mathcal{C}_n}=\frac{e^{\alpha_{n}l}}{l^{\frac{c}{8}}}$ with $c=\frac{1}{2}$. The natural expectation is that if we plug this formula in the 
definition of the Shannon information we get
\begin{eqnarray}\label{linear}
Sh(l)=\alpha l +\frac{c}{8}\ln l+...
\end{eqnarray}
where the dots are the subleading terms. The above formula  is consistent with \cite{AR2013}. However, one should be careful that although there are a lot of crystal configurations (polynomial number of them)
and "close" to crystal configurations that are connected to the central charge it is absolutely not clear what is going to happen in the scaling limit.
For example we repeated the calculations of \cite{Stephan2014} and realized that extraction of the coefficient of the $\ln$ in the above equation is 
indeed very difficult, see appendix. Here we show where one should look for the most important configurations. After a bit of inspection 
and numerical check, one can  see that the configuration with the highest probability is the $x=0$. Although the proof of the above statment
doesn't look 
straightforward one can understand it qualitatively  by starting from the ground state of the Ising model with $h\to\infty$ and approaching to the 
critical point $h=1$. 
The ground state of the Ising model with $h\to\infty$ 
is made of a configuration with all spins up. When we decrease the transverse magnetic field the other configurations start to appear
in the ground state. 
Although the amplitude of the configuration with all spins up decreases by decreasing $h$ it still remains always bigger than the other configurations.
Another way to look at this phenomena is by looking at the variation of the expectation value of the Hamiltonian
$\bar{H}=\bra{ C}H\ket{C}$ for different configurations $C$. It is easy to see that $\bar{H}$ is minimum for the configuration $C$ with all spins up. This
simply means that most probably when one construct the ground state of the Ising model using the variational techniques
this configuration playes the most important role. 
The least important configuration
is $x=1$ with the lowest probability. This can also be understood with the same heuristic 
argument as above. For every rank $k$ of the minors, as we discussed, we have $ \binom {l} {k}$ number of configurations which means that
for every $x$  for large $l$ we have $e^{f(x)l}$ with  $f(x)=-x\ln x-(1-x)\ln (1-x)$ number of configurations. It is obvious that the number of configurations
in every rank should be high enough to compensate the exponential decrease of probabilities. We realized that in every rank the configurations
$\textbf{a}$ has the lowest probabilities. One can again understand this fact using the variational argument. In this case it is
much better to make first the canonical transformation: $\sigma^x\to-\sigma^z$ and $\sigma^z\to\sigma^x$ in the Hamiltonian of the crtical
Ising model. Then one can simply argue that $\bar{H}$ is big if there are a lot of domain walls, i.e. $\bra{ C}\sigma_j^z\sigma_{j+1}^z\ket{ C}=-1$
in the system which is the case for the configurations
$\textbf{a}$.
Other important configurations are the configurations which divide the subsystem to two connected regions with in one part
all the spins are up and in the other part all the spins are down. These configurations are interesting because they have
the biggest probabilities among all the configurations corresponding to their minor rank. Note that in this case we have just one domain wall. 
It is not difficult to see that the probability of
all of these configurations decay exponentially  with the following coefficient $\alpha$:
\begin{eqnarray}\label{linear}
\alpha_{min}(x)=\frac{4C}{\pi}x+\ln 2-\frac{2C}{\pi},
\end{eqnarray}
where $C$ is the Catalan constant. In the two extreme points, we recover the previous results. We also checked the validity of the above formula numerically.
Having the biggest and smallest probabilities for every rank, we can now easily read the most important ranks. In Figure ~6 we depicted the
$\alpha_{max}$ and $\alpha_{min}$ for different configurations. We also depicted the graph of the number of configurations in every rank. 
The Figure clearly show that the configurations with $x>\frac{1}{2}$ can not have any significant contribution in the scaling limit because the 
number of configurations is not enough to compensate the exponential decay of the probabilities. A similar story seems to be  valid also for the values of
$x$ close to zero. The reason is that the number of configurations with small $x$ is such low that can not compensate
exponential decay of the probabilities in this region to have a significant contribution in the Shannon information. Just the region between the points that the two lines cut each other will most likely  survive in the scaling limit.
\begin{figure} [htbp!] \label{fig6}
\centering
\includegraphics[width=0.4\textwidth,angle =-90]{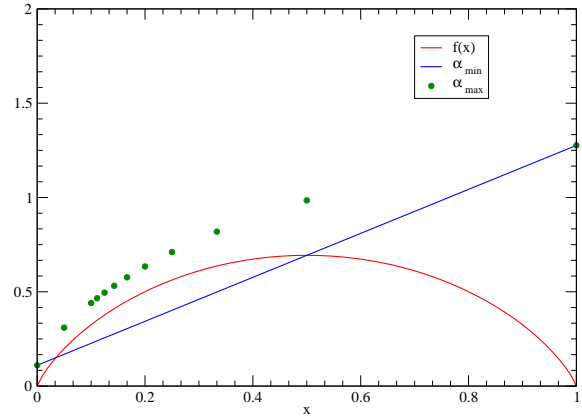}
\caption{(Color online) Values of $f(x)$, $\alpha_{min}$ and $\alpha_{max}$ with respect to $x$.
The red curve is the function $f(x)=-x\log x-(1-x)\ln (1-x)$ and the blue line is the linear function (\ref{linear}). 
The separated points are the $\alpha_{max}$ regarding the configurations discussed in the text.} 
\end{figure}
The numerical results indeed prove our expectation. In Figure ~7 we depicted the contribution of every rank $Sh_k(l)$ 
in Shannon information for two different sizes.
As it is quite clear 
the most important contributions come from $0<x<\frac{1}{2}$. The contribution of the configurations with $x>\frac{1}{2}$ is
exponentially small. This means that ignoring a lot of configurations will produce a very small amount of
error in the final result of Shannon information. To quantify this argument we calculated the amount of error in the evaluation of 
the Shannon information
if we just keep the configurations with ranks up to $x_m$. Suppose $Sh(l,x_m)$ is the contribution of the configurations with all ranks equal or 
smaller than $x_m$. Then the error of truncation can be calculated by $\mathcal{E}(x_m,l)=\frac{Sh(l,1)-Sh(l,x_m)}{Sh(l,1)}$. Interestingly we found that the logarithm
of the error function is a linear function of $l$, see Figure ~8. In other words
\begin{eqnarray}\label{error}
\ln\mathcal{E}(x_m,l)=-\lambda(x_m) l+\delta(x_m),
\end{eqnarray}
where $\lambda(x_m)$ is equal to zero and infinity for $x_m=0$ and $x_m=1$ respectively. $\lambda(x_m)$ for the other
values are shown in the inset of the Figure ~8. The above formula shows that one can calculate Shannon information with a controllable
accuracy by ignoring non-important configurations. Although the above truncation method help to calculate the Shannon information 
with good accuracy (especially the coefficient of the linear term $\alpha$)
it is still not good enough  to calculate the coefficient of the logarithm with controllable precision.
\begin{figure} [hthp!] \label{fig7}
\centering
\includegraphics[width=0.4\textwidth,angle =-90]{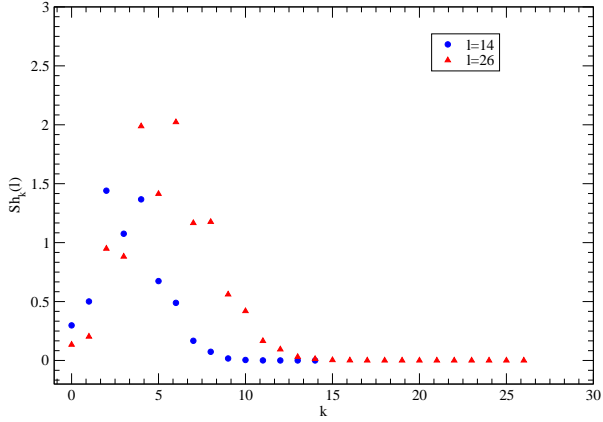}
\caption{(Color online) The contributions of different ranks $k$ in the Shannon information for two sizes $l=14$ and $26$.} 
\end{figure}

\begin{figure} [hthp!] \label{fig8}
\centering
\includegraphics[width=0.4\textwidth,angle =-90]{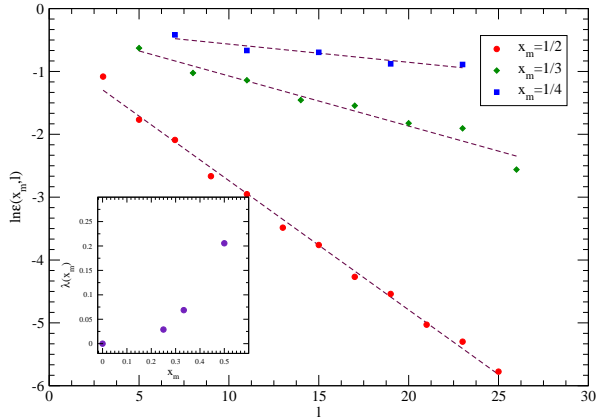}
\caption{(Color online) The error $\mathcal{E}(x_m,l)$ in the evaluation of Shannon information coming from the truncation at the rank $k=x_ml$. Inset: $-\lambda(x_m)$
with respect to $x_m$.} 
\end{figure}

\subsection{XX chain}

The Shannon information of the subsystem in the XX-chain is already discussed in \cite{Stephan2014} and based on numerical results
it is concluded that the equation (\ref{linear}) is valid with $\beta=\frac{1}{8}$ which is consistent with the conjecture in \cite{AR2013}.
Here we just comment on the contribution of different ranks which shows very different behavior from the transverse field Ising chain case. First of all,
as we discussed in the previous section when the external field is zero the only configurations that decay exponentially are those that
respect the half filling structure of the total system. The rest of the configurations scale like a Gaussian which simply indicates that
their contribution is very small in the Shannon information. This is simply because the number of these configurations scale just exponentially.
Based on this simple fact one can anticipate that the only configurations that can survive in the scaling limit are those with $k=\frac{l}{2}$.
Numerical results depicted in the Figure ~9 indeed support this idea. Although the $k=\frac{l}{2}$ is only one among $l$ possible minor ranks
the number of configurations with this rank is highest with respect to the others which can be one of the reasons that one can obtain
a good estimate for the coefficient of the logarithm in (\ref{linear}) with relatively modest sizes.

\begin{figure} [hthp!] \label{fig9}
\centering
\includegraphics[width=0.4\textwidth,angle =-90]{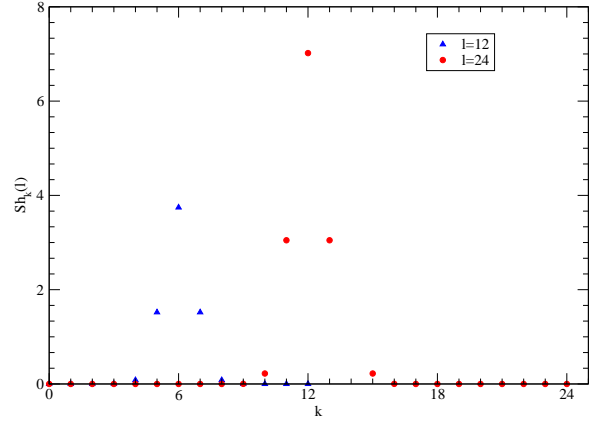}
\caption{(Color online) The contributions of different ranks $k$ in the Shannon information for two sizes $l=12$ and $24$ in the XX-chain.} 
\end{figure}

\section{Evolution of Shannon and  mutual information after global  quantum quench}

Inspired by experimental motivations, the field of quantum non-equilibrium systems has enjoyed a huge boost in the recent decade\cite{experimental}. One of the interesting
directions in this field is the study of information propagation after quantum quench, see for example \cite{CC-evolution, Plenio, LRbound, fidelity, Liu,MM}. 
Based on semiclassical arguments and also using Lieb-Robinson bound it is shown \cite{CC-evolution,LRbound}
that in one dimensional integrable system one can understand the evolution of entanglement entropy of a subsystem based on quasi-particle picture
\cite{CC-evolution}. 
The argument is as follows: after the quench, there is an extensive excess in energy which appears as quasiparticles that propagate
in time. The quasi-particles emitted from nearby points are entangled and they are responsible for the linear increase of the entanglement entropy
of a subsystem with respect to the rest.
In this section, we first study the time evolution of 
formation probabilities and subsequently Shannon and mutual information after a quantum quench. One can consider this section as a complement to
the other studies of information propagation after quantum quench.
To keep the discussion as simple as possible, we will concentrate on the most simple case of XX-chain or free fermions. 
Following \cite{PeschelEisler} consider the  Hamiltonian
\begin{eqnarray}\label{free fermion H}
H=-\frac{1}{2}\sum_{m=-\infty}^{+\infty}t_m(c_m^{\dagger}c_{m+1}+c_{m+1}^{\dagger}c_m).
\end{eqnarray}
\begin{figure} [hthp!] \label{fig10}
\centering
\includegraphics[width=0.4\textwidth,angle =-90]{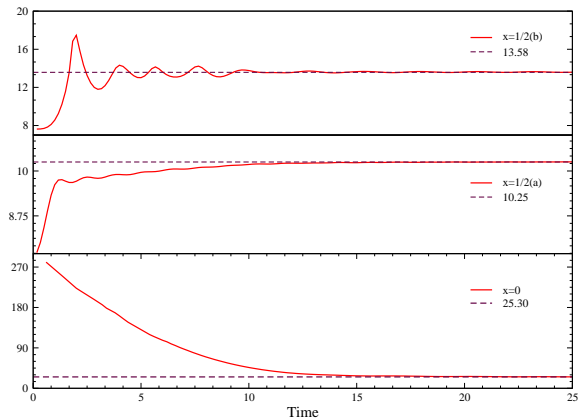}
\caption{(Color online) The evolution of logarithmic formation probability of different configurations with respect to time $t$ after quantum quench.
The size of the subsystem is taken $l=20$.} 
\end{figure}

The time evolution of the correlation functions in the half filling are given as 
\begin{eqnarray}\label{free fermion c matrix evolution}
C_{mn}(t)=i^{n-m}\sum_{jl} i^{j-l}J_{m-j}(t)J_{n-l}(t)C_{jl}(0),
\end{eqnarray}
where, $J$ is the Bessel function of the first kind. Here we consider  the dimerized initial conditions
with $t_{2m}=1$ and $t_{2m+1}=0$. The dimerized nature of the initial state will help later to consider
different possibilities for the initial Shannon mutual information. Then  at time zero we change the Hamiltonian to $t_m=1$ and let it evolve. The time evolution
of the correlation matrix is given by \cite{PeschelEisler}
\begin{widetext}
\begin{eqnarray}\label{correlation matrix evolution}
C_{mn}(t)&=&\frac{1}{2} \Big{(}\delta_{m,n}+\frac{1}{2}(\delta_{m+1,n}+\delta_{m-1,n})+e^{-i\frac{\pi}{2}(m+n)}\frac{i(m-n)}{2t}J_{m-n}(2t)\Big{)}.
\end{eqnarray}
\end{widetext}
\begin{figure} [hthp!] \label{fig11}
\centering
\includegraphics[width=0.4\textwidth,angle =-90]{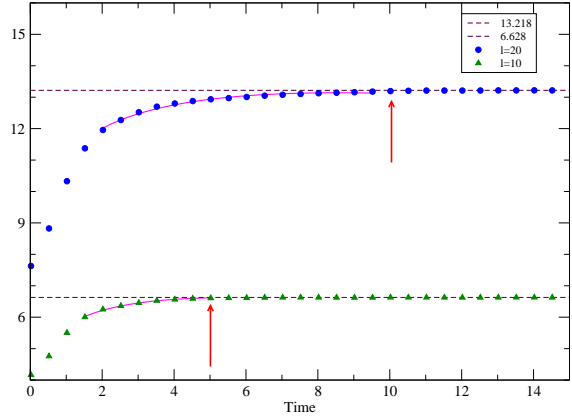}
\caption{(Color online) The evolution of Shannon information of a subsystem with different sizes  with respect to time $t$ after quantum quench. The
full lines are the equation (\ref{evolution}). The saturation points $t^*=\frac{l}{2}$ are marked by vertical arrows.
} 
\end{figure}
 To calculate the time evolution of the probability of different configurations one just needs to use the above
formula in (\ref{Formation probability XX}). The results for few configurations are shown in the Figure ~10. Of course 
since the sum of all the probabilities
should be equal to one some of the probabilities increase with time and some decrease. All the probabilities change rapidly
up to time $t^*\approx \frac{l}{2}$ and after that saturate.
One can also simply calculate the evolution of the Shannon information with the tools of previous sections. In Figure ~11
we depicted the evolution of Shannon information of a subsystem with respect to the time $t$. The numerical results show an increase 
in the Shannon information
 up to time $t^*\approx \frac{l}{2}$ and then saturation. This is  similar to what we usually have in the study of the time evolution of
 von Neumann entanglement entropy after quantum quench  \cite{CC-evolution}. However, one should be careful that in contrast 
to the von Neumann entropy the Shannon information
of the subsystem is not a measure of correlation between the two subsystems. In addition, the increase in the Shannon information 
of the subsystem is not linear as the evolution of the von Neumann entanglement entropy. Our numerical results indicate that apart from a 
small regime at the beginning the Shannon information increases as
\begin{eqnarray}\label{evolution}
Sh(l)=alt^b-dt \hspace{1cm}t<t^*
\end{eqnarray}
where $b\approx0.15(2)$ and $a$ and $d$ are positive $l$ independent quantities.

\begin{figure} [hthp!] \label{fig12}
\centering
\includegraphics[width=0.4\textwidth,angle =-90]{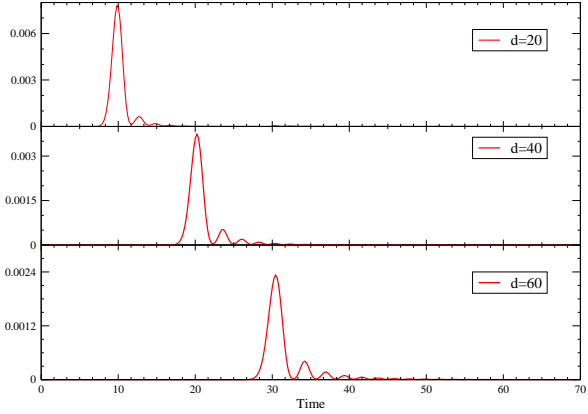}
\caption{(Color online) The mutual information between a pair of dimers located at   distance $d$ with respect to time.} 
\end{figure}

\begin{figure} [hthp!] \label{fig13}
\centering
\includegraphics[width=0.4\textwidth,angle =-90]{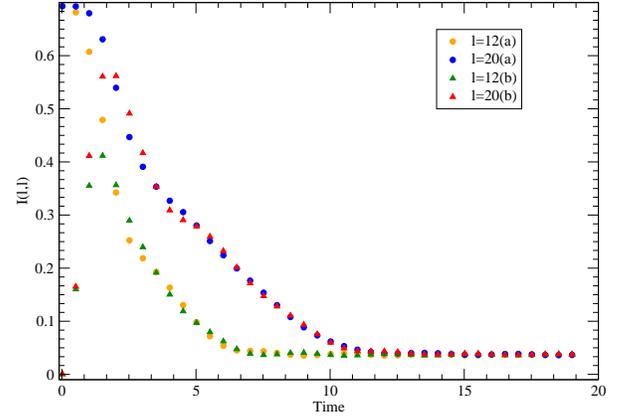}
\caption{(Color online) The evolution of mutual information between two adjacent subsystems in two different cases:
when at the boundary between
the two subsystems there is a dimer, case $a$ and when there is no dimer, case $b$. In the first case
the mutual information starts from a non-zero value but in the second case it starts from zero. } 
\end{figure}
To study the time evolution of correlations, it is much better to
study another quantity, Shannon mutual information of two subsystems. To investigate this quantity we first studied the time evolution
of Shannon mutual information of   a couple of dimers located far from each other. The results depicted in
the Figure ~12 show that the Shannon mutual information of the dimers are zero up to time $t^*\approx\frac{l}{2}$ and after that increases rapidly
and then again decays slowly. This picture is  consistent with the quasiparticle picture. The two regions are not correlated up to time that the
quasiparticles emitted from the middle point reach each dimer\cite{footnote3}. However, the similarity between
the evolution of the von Neumann entropy and mutual Shannon
 information ends here. To elaborate on that we consider the mutual information of two adjacent regions with sizes $\frac{l}{2}$. 
Because
of the dimerized nature of the initial state there are two possibilities for choosing the subsystems: at time zero at the boundary between the two subsystems
there can be a dimer or not. In the first case  at time zero the Shannon mutual information between the two subsystems is not zero but
in the second case it is  zero. In the second case naturally one expects an overall increase in the mutual information but
in the first case a priory it is not clear that the mutual information should increase or decrease. In the Figure ~13 we have depicted
the results of the numerics for the two adjacent subsystems for different sizes. The numerical results show that for the uncorrelated 
initial conditions the Shannon mutual information first increases rapidly and then it decays and finally saturates
at time $t^*=\frac{l}{2}$. In the correlated case, we have overall decay in the mutual information and finally the saturation
again at time $t^*=\frac{l}{2}$. This behavior is very different from the quantum mutual information of the same regions which for the
considered initial states first increases
linearly and then saturates at time $t^*=\frac{l}{2}$. The interesting phenomena is that after a short initial regime that the Shannon
mutual information is initial state dependent the system enters to a regime that this quantity is completely independent of initial state and it
decreases "almost" linearly and then saturates. This can be also easily seen from the equation
(\ref{evolution}), where we can simply drive 
\begin{eqnarray}\label{evolution MI}
I(l,l)=-dt \hspace{1cm}t<t^*.
\end{eqnarray}
The saturation regime is independent of the size of the subsystem, this is simply because
in the equlibrium regime the Shannon mutual information follows the area-law \cite{Wolf2008} and so it is independent of the volume of the subsystems.

\section{Conclusions}

In this paper, we employed Grassmann numbers to write the probability of occurrence of 
different configurations in free fermion systems with respect to the minors of a particular matrix. The formula gives a very efficient method
to study the scaling properties of logarithmic formation probabilities in the critical XY-chain. In particular, we showed that the logarithmic
formation probabilities of crystal configurations are given by the CFT formulas for the critical transverse field Ising model.
This is checked by studying the probabilities in the infinite and finite (periodic and open boundary conditions) chain. In the case of critical XX-chain 
which has a $U(1)$ symmetry just the configurations with $x=\frac{n_f}{\pi}$ follow the CFT formulas. The rest of the configurations
decay like a Gaussian and do not show much universal behavior.  We also studied the Shannon information of a subsystem in the transverse
field Ising model and XX-chain. In particular, for the Ising model, we showed that in the scaling limit just the configurations with a high number of
up spins contribute to the scaling of the Shannon information. In principle, if one considers all the configurations, with our method one can not calculate
the Shannon information with classical computers for sizes bigger than $l=40$ in a reasonable time.
However, if one admits a controllable error in the calculation of Shannon information it is possible to hire the results
of section V to go to higher sizes. It would be very nice to extend this aspect of our calculations further to calculate 
the universal quantities in the Shannon information with higher accuracy. For example, one interesting direction can be finding an explicit
formula for the sum of different powers of principal minors of a matrix. This kind of formulas can be very useful to calculate analytically or
numerically the R\'enyi entropy of the subsystem. 

Finally, we also studied the evolution of formation probabilities
after quantum quench in free fermion system. In this case, we prepared the system in the dimer configuration and then we let it evolve with
homogeneous Hamiltonian. The evolution of Shannon information of a susbsytem shows a very similar
behavior as the evolution of entanglement entropy after a quantum quench. Especially our calculations show that the saturation of the Shannon
information of the subsystem occurs at the same time as the entanglement entropy. This is probably not surprising because the $t=t^*$ is also
the time that the reduced density operator saturates. 

It will be very nice to  extend our calculations in few other directions. One direction can be investigating the evolution of mutual information after
local quantum quenches as it is done extensively in the studies of the entanglement entropy \cite{CClocal,MP,Karevski}. The other interesting direction can be
calculating the same quantities in other bases, especially those bases that do not have any direct connection to CFT.

\paragraph*{Acknowledgments.}

We thank M Fagotti and F Alcaraz for many useful discussions. The work of MAR was supported in part by
CNPq. KN acknowledges support from the National Science Foundation under grant number PHY-1314295. 
MAR also thanks ICTP for hospitality during a period which part of this work was completed.

\textbf{Appendix A: Shannon information for critical transverse-field Ising model } 

In this Appendix, we will provide more details regarding the Shannon information of transverse critical Ising chain and XX chain.
The data regarding Shannon information for a subsystem with length $l$ up to $l=39$ is listed in the Table A1. Having
the data, we checked many different functions with different parameters to study the coefficient of the logarithm. Needless to say 
increasing the possible parameters can make a difference in the final result. In \cite{Stephan2014} the results  fitted to 
\begin{eqnarray}\label{Stephan}
Sh(l)=\alpha l+\beta \ln l +\sum_{n=1}^5 \frac{b_n}{l^n}+\delta
\end{eqnarray}
show that  the best value is $\beta=0.060$. This can be also checked using the data provided in the Table A1. It is worth mentioning that one can also
get reasonable results using the data up to $l=40$ for some formation probabilities (not all) if we consider extra terms $\sum_{n=1}^5 \frac{b_n}{l^n}$
in the fitting procedure. Although we found that the equation (\ref{Stephan}) is the most stable fit with the least standard deviation based
on our results in the main text we found it is hard to exclude the term $\frac{\ln l}{l}$ because it is present in all the configurations studied there.
If one includes this term and does not add the terms $\sum_{n=1}^5 \frac{b_n}{l^n}$ the $\beta$ coefficient will be $0.0617$. If one keeps all the terms
$\sum_{n=1}^5 \frac{b_n}{l^n}$ the result will be $\beta=0.060$. The final conclusion is that as far as one justifies the presence of the terms
$\sum_{n=1}^5 \frac{b_n}{l^n}$ in the Shannon information formula the best value for $\beta$ with the current
 available data is $0.060$.
\setcounter{table}{0}
\makeatletter 
\renewcommand{\thetable}{A\arabic{table}}
\begin{table}[hthp!]
\centering
{\begin{tabular}{|l|c|c|c|c|}
  \hline
l           &      Shannon           &  l           &      Shannon\\
  \hline
1           &   0.473946633733778    &  21          &   9.094267377324401     \\
2           &   0.925441055292197    &  22          &   9.520258511384927     \\
3           &   1.367970612016317    &  23          &   9.946131954351737     \\
4           &   1.805854593071358    &  24          &   10.37189747498959     \\
5           &   2.240889870728481    &  25          &   10.79756367448224     \\
6           &   2.674003797245196    &  26          &   11.22313816533366     \\
7           &   3.105734740754158    &  27          &   11.64862771729621     \\   
8           &   3.536422963908594    &  28          &   12.07403837729498     \\ 
9           &   3.966297046625437    &  29          &   12.49937556879184     \\ 
10          &   4.395517906953372    &  30          &   12.92464417475445     \\     
11          &   4.824203084194648    &  31          &   13.34984860684562     \\    
12          &   5.252441034545332    &  32          &   13.77499286454422     \\     
13          &   5.473946633733777    &  33          &   14.21026702317442     \\
14          &   6.107833679024358    &  34          &   14.62511508430369     \\       
15          &   6.535085171703405    &  35          &   15.05009939651494     \\     
16          &   6.962089515106671    &  36          &   15.47503630258492     \\   
17          &   7.388875612253789    &  37          &   15.89992835887542     \\      
18          &   7.815467577831834    &  38          &   16.32477792018708     \\      
19          &   8.241885740227190    &  39          &   16.74960160654153     \\     
20          &   8.668147394540807    &              &                         \\  

   \hline
\end{tabular}}
\caption{Shannon information calculated for sizes $l=1,2,...,39$.} 
\end{table}


\begin{thebibliography}{99}
\bibitem{footnote1}Similar oscillating behavior
exist also in non-critical regime with high transverse magnetic field. In this case the ffect is already understood based on
superconducting correlations of real fermions in \cite{Franchini}.
\bibitem{footnote2} Although not shown in the table it is worth mentioning that for those configurations
   that depending on the parity of the number of down spins we have oscillating behavior the extracted value of
   $\nu$ is not exactly the same for those cases that $m$ is odd or even. However we conjecture
   that this is just a finite size effect and the formula (\ref{main formula2}) is correct.
\bibitem{footnote3} Recently in \cite{MM} it has been pointed out that
because of the presence of integrals of motions instead of two quasiparticles one needs to consider four of them. It is not clear for us how this can change
the results of our investigation.
\end{thebibliography}
\end{document}